\documentclass[12pt,preprint]{aastex} 
\slugcomment{} 

\newcommand\kms{\ifmmode {\rm km\ s}^{-1} \else km s$^{-1}$\fi}
\newcommand\eflux{\ifmmode {\rm ergs\ s}^{-1}\;{\rm cm}^{-2} \else  
	ergs s$^{-1}$ cm$^{-2}$\fi}  
\newcommand\phflux{\ifmmode {\rm photons\ s}^{-1}\;{\rm cm}^{-2} 
	\else  	photons s$^{-1}$ cm$^{-2}$\fi}  
\newcommand\ergsec{\ifmmode {\rm ergs\ s}^{-1} \else  
	ergs s$^{-1}$\fi}
\newcommand\Msun{\ifmmode M_{\odot} \else $M_{\odot}$\fi} 
\newcommand\gtsim{\raisebox{-.5ex}{$\;\stackrel{>}{\sim}\;$}}

\shorttitle{The SED of Ton~S180} 
\shortauthors{Turner et al.\ 2001} 
 
\begin{document}
\title{The Spectral Energy Distribution of the Seyfert galaxy Ton~S180\footnote{Based 
	(in part) on observations made with the Danish 1.5m 
	telescope at ESO, La Silla, Chile}}
\author{T.\ J.\ Turner\altaffilmark{1,2}, 
	P.\ Romano\altaffilmark{3},  
	S.\ B.\ Kraemer\altaffilmark{4, 5}, 
	I.\ M.\  George\altaffilmark{1,2},
	T.\ Yaqoob\altaffilmark{1,6}, 
	D.\ M.\ Crenshaw\altaffilmark{7},  
	J.\ Storm\altaffilmark{8}, D.\ Alloin\altaffilmark{9}, 
	D.\ Lazzaro\altaffilmark{10}, L.\ Da Silva\altaffilmark{10}, 
	J.\ D.\ Pritchard\altaffilmark{9}, 
	G.\ Kriss\altaffilmark{6,11}, W.\ Zheng\altaffilmark{6}, 
	S. Mathur\altaffilmark{3}, 
 	J.\ Wang\altaffilmark{6},  
P.\ Dobbie\altaffilmark{12}, N.\ R.\ Collins\altaffilmark{4,5,13}}

\altaffiltext{1}{Joint Center for Astrophysics, Physics Dept., University of Maryland
	Baltimore County, 1000 Hilltop Circle, Baltimore, MD 21250} 
\altaffiltext{2} {Laboratory for High Energy Astrophysics, Code 662, 
	NASA/GSFC, Greenbelt, MD 20771}
\altaffiltext{3} {Department of Astronomy, The Ohio State University, 140 West Avenue, 
	Columbus, OH 43210} 
\altaffiltext{4} {Laboratory for Astronomy and Solar Phyiscs, Code 661, 
	NASA/GSFC, Greenbelt, MD 20771}
\altaffiltext{5}{Institute for Astrophysics and Computational Sciences,
        The Catholic University of America
        Washington, DC   20064}
\altaffiltext{6}{Center for Astrophysical Sciences, 
	Department of Physics and Astronomy, The Johns Hopkins 
	University, Baltimore, MD 21218}
\altaffiltext{7}{Department of Physics and Astronomy, 
	Georgia State University, Atlanta, GA 30303}
\altaffiltext{8}{AIP, An der Sternwarte 16, D-14482 Potsdam, Germany}
\altaffiltext{9}{European Southern Observatory, Casilla 19001, 
	Santiago, Chile }
\altaffiltext{10}{Observatorio Nacional, Departamento de 
	Astronomia, Rua Gen.    Jose Cristino, 77, 20921-400, Rio de Janeiro, 
   	Brazil Observatorio Nacional, Rio de Janeiro, Brazil }
\altaffiltext{11}{Space Telescope Science Institute, 3700 San Martin 
	Drive, Baltimore, MD 21218}
\altaffiltext{12}{Astronomy Group, Leicester University, 
	Leicester LE1 7RH, UK}
\altaffiltext{13}{Science Systems and Applications, Inc., 
	5900 Princess Garden Parkway Suite 300, Lanham, MD  20706
}
	\begin{abstract}
We present spectral results from a multi-satellite, broad-band 
campaign on the Narrow-line Seyfert 1 galaxy \objectname[PHL 912]{Ton~S180} 
performed at the end of 1999. We discuss 
the spectral-energy distribution of the source, combining 
simultaneous {\it Chandra}, {\it ASCA} and {\it EUVE} 
data with contemporaneous {\it FUSE}, {\it HST}, and 
ground-based optical and infra-red data. 
The resulting SED shows that most of the energy is emitted in the 
10 -- 100 eV regime, which must be dominated by the primary energy source. 
No spectral turnover is evident in the UV regime. This, the strong 
soft X-ray emission, and the 
overall shape of the SED indicate that emission from the accretion disk peaks 
between 15 and 100 eV.
High resolution {\it FUSE} spectra showing UV absorption due to O{\sc vi} 
and the lack of detectable X-ray absorption in the {\it Chandra} 
spectrum demonstrate the presence of a low column density of highly 
ionized gas along our line-of-sight.

The highly-ionized state of the circumnuclear gas 
is most likely linked to the high luminosity and steep spectrum 
of the active nucleus. 
Given the strong ionizing flux in Ton~S180, 
it is possible that the clouds within a few tens of light days
of the central source are too highly ionized to produce much line 
emission. Thus the 
narrow width of the emission lines in Ton~S180 is due to the 
emission arising from large radii.

	\end{abstract}


	\section{Introduction}

A longstanding suggestion has been that there is a so-called ``big-blue-bump''
(BBB) of continuum emission, peaking in the unseen  X-ray to UV (XUV) regime, 
perhaps originating in the accretion disk. 
There are indications of the low energy tail of this component in 
the UV spectra of Seyfert 1 galaxies 
\citep[][]{Shields78,MalkanSa82}. 
The peak energy of the disk emission is predicted to be 
dependent on the accretion rate \citep[][]{MFR93b}. Thus 
the spectral-energy-distribution (SED) of an AGN 
provides crucial information about accretion rates and conditions 
close to the disk.
However, determination of the XUV continuum in 
AGN has been extremely difficult because of the severe attenuation of
photons of these energies by even small amounts of Galactic material along
the line-of-sight. Some indication of the strength of 
the unseen continuum has been inferred from the strengths of emission lines
such as \ion{He}2$ \lambda1640$ \citep[e.g.][]{MathewsF87}. 
\citet{Zhengea97} have suggested the form of the
unseen XUV spectrum is $f_v \propto v^{-\alpha}$ with $\alpha=2$ 
between the Lyman limit (at 912
\AA) and $\sim 0.5 $\,keV. \citet{Laorea97} combine this with a mean soft
X-ray spectrum, based upon {\it ROSAT} observations of quasars, to dispute
the existence of a large XUV bump. \citet{Telfer02} extend the work of 
\citet{Zhengea97} by including more data in the extreme UV band; those 
authors find the data to be adequately represented by a simple power-law 
with $\alpha=1.76$ between 500 and 1200 \AA . \citet[][]{KoristaFB97} have
discussed the problem that extrapolating the known soft X-ray spectrum of
AGN, there appear to be too few 54.4 eV photons to account for the strength
of the observed \ion{He}{2} lines. They consider the possibility that the
broad-line clouds see a harder continuum than the observer does, or that
the XUV spectrum has a double-peaked shape. 

Narrow-line Seyfert 1 galaxies (NLSy1s) 
were first classified on the basis of their unusual 
optical properties, most notably, lying at the lower end of the 
distribution of line widths, for permitted optical lines. 
Specifically, widths of 
H$\beta$ FWHM $<2000$\,\kms{} have been taken 
as the defining quality of NLSy1s, although it is widely 
acknowledged that this value is arbitrary and there is a continuous 
range of broad-band properties across the Seyfert population.  
Nevertheless, the distinction is useful, as an 
object's place in the line-width distribution of the Seyfert population 
is an indicator of other properties, especially the X-ray properties. 
NLSy1s appear to have systematically 
different, or very extreme X-ray attributes compared to the rest of the
Seyfert population, which we will henceforth refer to as broad-line-Seyfert
1s (BLSy1s). For example, examination of X-ray properties across the
Seyfert population reveals that FWHM H$\beta$ is strongly anti-correlated
with the excess variance $\sigma^{2}_{RMS}$  
(\citealt{TGNT99}; 
where $ \sigma^{2}_{\rm RMS}=\frac{1}{N\mu^{2}}\sum_{i=1}^{N} [(X_{\rm
i}-\mu)^2-\sigma_{\rm i}^{2}] $ for a light curve 
with $N$ points of counts $X_{\rm i}$ and 
unweighted mean  $\mu$) 
and anti-correlated with  X-ray
index, in the sense that NLSy1s are most variable and have steeper spectra
\citep{BBF96, BME97}. These correlations suggest a
fundamental difference between BLSy1s and NLSy1s, the 
latter being thought to represent the low mass and/or high accretion-rate 
systems \citep[e.g.][]{PDO95}. 

For black holes operating near $L_{\rm Edd}$, 
the accretion disk surface is predicted to be highly ionized, thus the disk
spectrum is expected to peak at higher energies in NLSy1s than in 
BLSy1s. In the soft X-ray regime the NLSy1s 
Ton~S180 and Arakelian~564 (Akn~564) show a hump of emission close to 1\,keV 
\citep[e.g.][]{TGN98,Vaughan99a,Turnerea01,Akn564I}, which
could be the signature of a hot disk. \citet{Fioreea98} also suggested an
ionized disk as the origin of a soft X-ray component in PG $1244+026$.
Interestingly, at least nine AGN are known to possess a soft hump 
with characteristic shape as described for Ton~S180. These AGN 
 are all NLSy1s: PG~1244$+026$ \citep{Fioreea98,Gea00};
IRAS~13224-3809 \citep{Vaughan99b}; Akn~564 \citep{BFNRB94,TGN99};
NGC~4051 \citep{Collea01}; PG~1404$+226$, PG~1440$+356$ (Mkn 478), and
PG~1211$+143$ \citep{Gea00}; Ton~S180 \citep{TGN98, Turnerea01}; 
1H 0707-495 \citep{Boll01}.
Further evidence for an ionized disk comes from the
observation of Fe K$\alpha$ emission from ionized gas in numerous
NLSy1s \citep{Comastriea98,TGN98,TGN99,Vaughan99b,Comastriea01,BIF01}.

Thus examination of the SED of a NLSy1, and comparison 
with that obtained for BLSy1s should offer insight into the 
relative accretion rates across the Seyfert population. 
To this end, we undertook a multi-wavelength campaign to 
obtain a broad-band spectral-energy-distribution of Ton~S180. 

\section{Ton~S180}

We selected Ton~S180 for study as it is bright in the soft 
X-ray regime and has low line-of-sight and intrinsic 
extinction, allowing  a view of the bare continuum form. 
Ton~S180  (PHL 912, z=0.06198; \citealt{Wisotzkiea95}) has a low 
 Galactic column density along the line-of-sight,  
$N_{\rm H}=1.55^{+0.27}_{-0.13} \times 10^{20}$ cm$^{-2}$ \citep{dicklock90}.
The uncertainty on the column density represents the maximum scatter 
of values of column density within 
a 1 degree cone centered on Ton~S180. 
The observed flux (i.e.\ no correction for extinction) is 
$F_{0.5-2 {\rm \,\,keV}} \sim 1.1 \times 10^{-11}$ \eflux{} \citep{TGN98}. 
The source is at the extreme end of the Seyfert range of line widths 
with FWHM H$\alpha$ and H$\beta \sim 900$\,\kms, making 
it a good choice for isolating the fundamental parameter which determines 
the classification of a Seyfert galaxy. 

{\it BeppoSAX} \citep{Comastriea98} and {\it ASCA} \citep{TGN98}
data from Ton~S180 indicated a steep 
spectrum in the 2--10\,keV band, with $\alpha= 1.5$; both datasets 
also showed an Fe K$\alpha$ emission line peaked at 
a rest-energy $\sim 7$\,keV, indicating that 
the circumnuclear material is strongly ionized. 
{\it ASCA} data confirm the complexity of the  soft X-ray spectrum 
first noted in {\it ROSAT}\ PSPC data \citep{Finkea97}. 
A {\it Chandra} Low Energy Transmission Grating (LETG) 
observation has recently revealed the soft hump
component to be a smooth continuum or extremely broadened 
reprocessed component \citep{Turnerea01}. 

In this paper we use 
energy index $\alpha$ for quantification of spectral indices, 
defined such that the flux density  
$F(E) \propto E^{-\alpha}$ at energy $E$. 
A log of the observations performed in support of the campaign is presented
in Table~\ref{obslogs}. All the data were reduced using standard techniques as
outlined below.

\section{ASCA}
\subsection{Data Reduction}
\begin{figure}	 
	\epsscale{0.9} 
	\vspace{-1.6truecm}
	\plotone{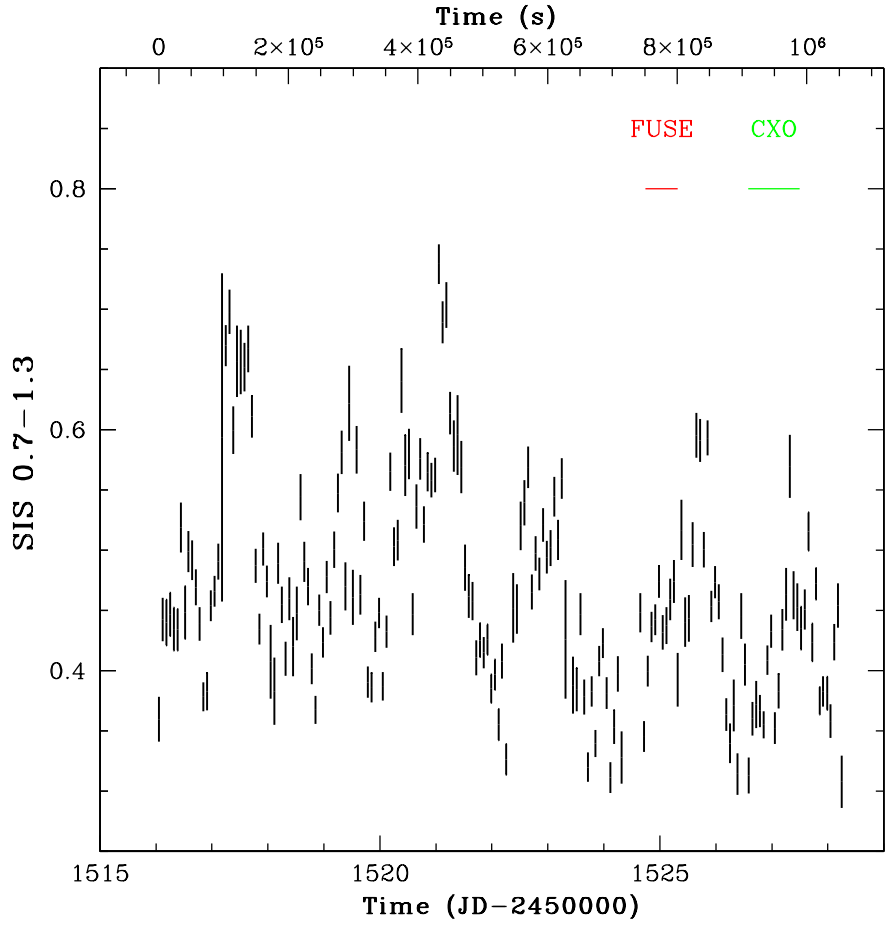}
	\caption[{\it FUSE} and {\it Chandra} versus ASCA]
{The 12-day {\it ASCA} light curve from \citet{Romanoea02}. 
The periods covered by  {\it FUSE} and {\it Chandra} observations are noted.
{\it EUVE} data were extracted to be simultaneous with the {\it Chandra} 
observation. 
\label{fig:lc} 
}
\end{figure}

The {\it ASCA} satellite carries four focal-plane detectors, 
two CCDs (the Solid-state Imaging Spectrometers or SISs, 
covering 0.4-10.0 keV) and 
two Gas Imaging Spectrometers (GISs, covering 0.7-10.0 keV), 
all are operated simultaneously. {\it ASCA} observed Ton~S180 for a baseline 
of 12 days, $\sim 1$Ms,  1999 December 3 -- 15. Those data were 
reduced using the methods and screening 
criteria of the {\it Tartarus} \citep{Turnerea99b} 
database. As reported by \citet{Romanoea02} these screening criteria 
resulted in an effective exposure of 
405\,ks in the GISs, 338\,ks and 368\,ks  in SIS0 and SIS1, respectively. 
The mean SIS0 count rate was $0.586\pm0.001$ ct s$^{-1}$. 
The data were reduced using the calibration file 
\verb+sisph2pi_290301.fits+ and the degradation of the low-energy response 
of the SISs was compensated for by the time-dependent absorption term 
detailed by \citet{Yaqoob}\footnote{see http://heasarc.gsfc.nasa.gov/docs/asca/calibration/nhparam.html}. 
\citet{Romanoea02} utilized corrections of 
$N_H=6.9 \times 10^{20} {\rm cm^{-2}}$ and 
$N_H=1.0 \times 10^{21} {\rm cm^{-2}}$ for SIS0 and SIS1, respectively. 
The results presented here are primarily based on 
the analysis of the time-averaged spectra obtained during the 
simultaneous {\it ASCA}--{\it Chandra} observations. 
This means that while all of the {\it Chandra} data were used, only 
a subset of the {\it ASCA} data were utilized. 
Figure~\ref{fig:lc} shows the periods covered by the {\it FUSE} and 
{\it Chandra} observations, with respect to the entire 12-day 
observation by {\it ASCA}, allowing us to see where these new 
datasets lie compared to the recent flux history of the source. 
Figure~\ref{fig:axafasca_lc256}, shows in detail the {\it Chandra} light curve 
with part of the overlapping {\it ASCA} dataset. 
In this paper, we use only the mean {\it ASCA} spectrum  
from the period simultaneous with the {\it Chandra} observation,  
i.e. within JD  2451526.576 -- 2451527.498. 

\subsection{ASCA Results}

As is evident from 
Figure~\ref{fig:axafasca_lc256}, Ton~S180 exhibited significant 
changes in flux. \citet{Romanoea02} present a detailed analysis of 
spectral variability 
over the full 12-day observation. 
\citet{Romanoea02} find 
the continuum fit to the mean spectrum
to yield $\alpha = 1.44 \pm 0.02$. A  
strong excess of emission is observed below 2\,keV, and this soft component 
varies in strength down to the minimum timescale determinable via spectral 
analysis, $\sim 1$ day. The variations in hump strength are correlated with
the photon index and the 2-10 keV flux, consistent with disk-corona models
\citep{Romanoea02}. 
The softness ratio shows rapid variability on timescales $< 1000$ s, 
indicating either a breakdown of the correlation between soft hump and 
power-law fluxes on such short timescales, or rapid variations in the 
photon index. 
\citet{Romanoea02} also find a broad Fe K$\alpha$ line with narrow peak at a
rest-energy 6.8 keV, indicating an origin in ionized material.

Analysis of the spectrum 
acquired simultaneously with {\it Chandra} data 
reveals $\alpha=1.44\pm0.07$ (in agreement with the mean for the full 
dataset). The soft component shows an 
equivalent width EW $=63^{+71}_{-50}$\,eV when parameterized using a Gaussian 
model. As \citet{Romanoea02} found no evidence for variability in the flux 
or equivalent width of the Fe K$\alpha$ line we do not fit for the 
Fe K$\alpha$ parameters here (and exclude the 5.0-7.5 keV data 
when fitting for continuum slopes).

\section{Chandra}
\subsection{ACIS/LETG Data Reduction}

The {\it Chandra} data were reprocessed using 
calibration files from {\tt CALBD v2.6}. the data were then  
screened to remove bad pixels, columns and 
events with detector `grades' {\it not} equal to 0, 2, 3, 4, or 6.
Periods of high background were also excluded.
Such screening resulted in an on-source exposure of $\sim$75~ks.
The 1$^{st}$-order spectra were extracted from the screened event file
and appropriate ancillary response files constructed using 
{\tt CIAO v2.1}. 
Previous analysis of these data \citep{Turnerea01} 
was limited to use of data above $\sim 0.4$ keV, due to 
unacceptable uncertainty in the ACIS/LETG calibration around the 
C-edge. However, the quantum efficiency (QE) file 
\verb+acisD1997-04-17qeN0004.fits+, released 
2001-06-07, was utilized in the analysis presented here, this 
improves the fit to ACIS/LETG observations 
of calibration sources, in the C-edge regime. 
Utilization of the new quantum efficiency file 
allows examination of data down to 0.2 keV, the lowest 
energy available from LETG data with ACIS in the focal plane. 
The entire {\it Chandra} baseline is utilized in this analysis, as the 
{\it ASCA} observation overlaps the {\it Chandra} observation completely 
(Figure~\ref{fig:axafasca_lc256}). 

\subsection{LETG Results}

\begin{figure} 
	\epsscale{0.8} 
	\plotone{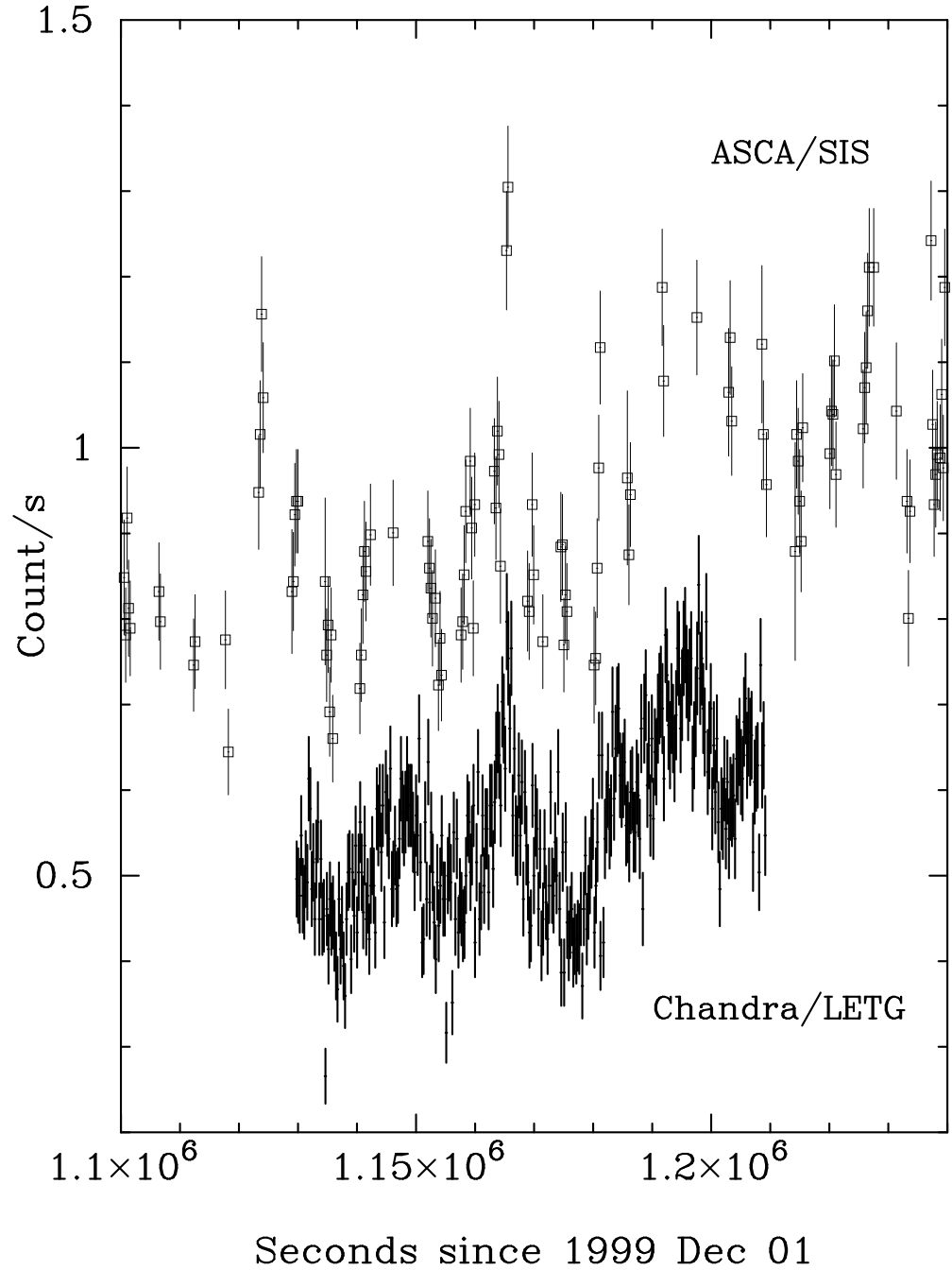} 
	\caption[]{The light curve of the combined 
$\pm$1$^{st}$ order {\it Chandra}/LETG data (0.2-10.0\,keV) using 256\,s bins.
The corresponding portion of the {\it ASCA} SIS light curve 
is overlaid and denoted by the open symbols. 
\label{fig:axafasca_lc256}
} 
\end{figure}

In Figure~\ref{fig:axafasca_lc256} we show the light curve
obtained from the $\pm$1$^{st}$-order
{\it Chandra}/LETG data.
The portion of the {\it ASCA} SIS light curve
(from Figure~\ref{fig:axafasca_lc256}) is overlaid 
for direct comparison.
As might be expected, there is good agreement between the 
light curves from the two instruments.

\begin{figure}	 
	\epsscale{0.9} 
	\plotone{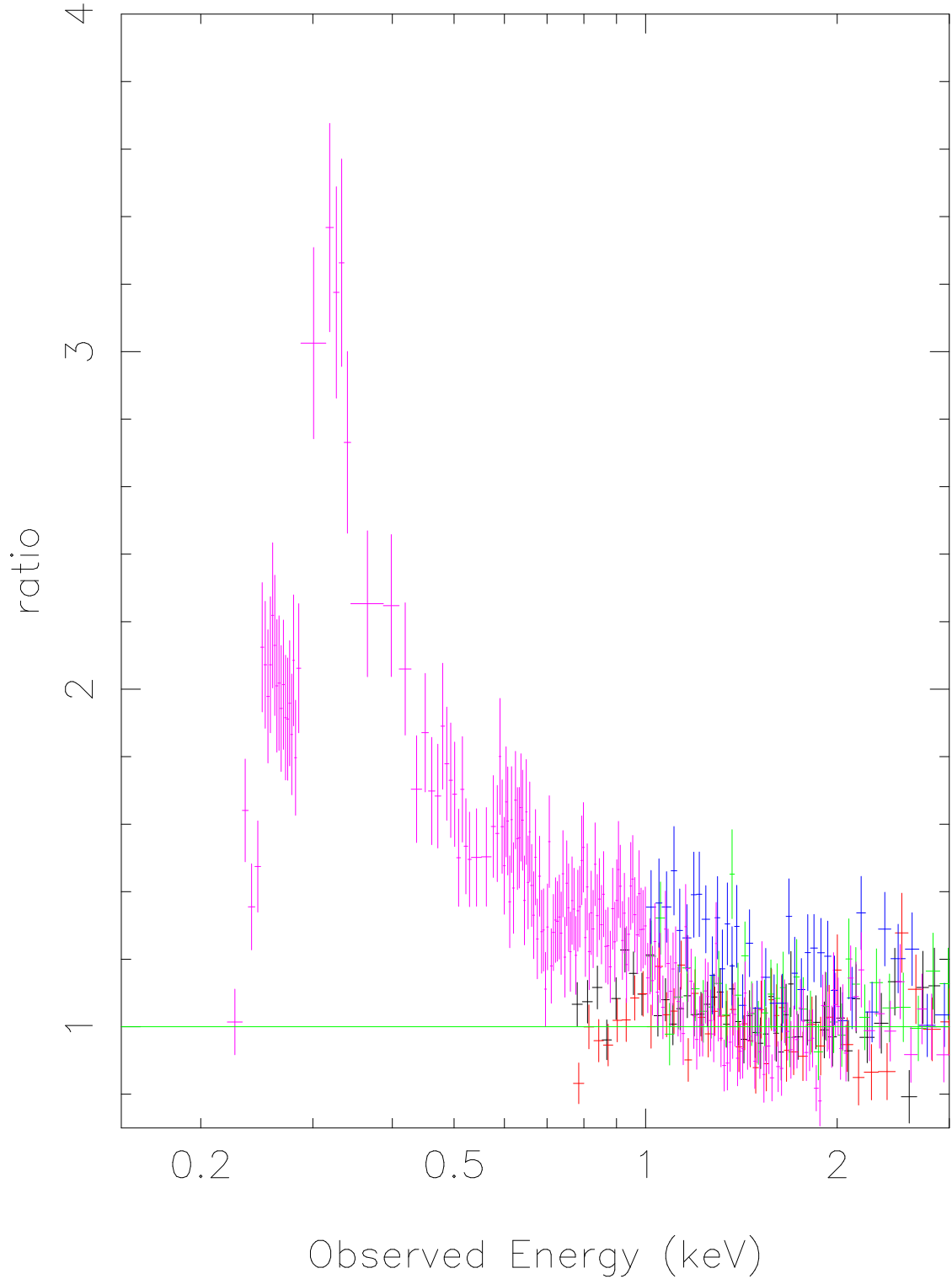} 
	\caption[]{The LETG (magenta) and {\it ASCA} (red and black are SIS,
blue and green are GIS) data/model ratio compared to 
the $\alpha=1.44$ power-law. The soft excess shows a 
shape which may be partially 
due to residual calibration uncertainties below 0.3 keV.
\label{fig:excess_shape}
}
\end{figure}

\citet{Turnerea01} present the first order LETG spectra of Ton~S180, 
finding no strong spectral features and concluding that the 
excess soft X-ray emission discovered using
{\it ASCA} \citep{TGN98} must be primarily due to a
previously-unknown {\it continuum} component or very broadened 
reprocessed component.  
\citet{Turnerea01} note the lack of strong absorption features in
the X-ray spectrum of Ton~S180, in contrast to
results from many Seyfert 1 galaxies
(e.g.\ the BLSy1s NGC~5548, \citealt{Kaastraea00}, and NGC~3783 \citealt{K2000}, 
and the NLSy1 NGC~4051, \citealt{Collea01}).
Analysis using the new QE file reveals an improvement to the agreement between
{\it ASCA} and LETG data in the 1-2 keV regime (c.f.\ results presented
by \citealt{Turnerea01}). We also re-examined the shape of the soft excess.
The extrapolation of the hard-band power-law
to soft X-ray energies reconfirms the presence of
excess soft emission as expected, with a sharp turnover of
the data below $\sim 0.3$\,keV ($\sim 7 \times 10^{16}$\,Hz).
Figure~\ref{fig:excess_shape} shows
the form of the soft component. The turnover is sharper than that
expected due to absorption by edges in neutral or ionized gas.
In any case, if there were such a deep absorption edge in the
X-ray regime, strong spectral features would be expected in other
parts of the spectrum, which are not observed.
The soft hump in the data was modeled using the {\sc xspec}
{\sc diskbb} model (\citealt{mitsuda84,mak86}).
However, although
{\sc diskbb} has some intrinsic spectral curvature, this
was not able to account for the shape of the LETG data.
A fit to data above 0.3 keV yielded a best-fit
temperature of $\sim 98$ eV at the inner edge of 
the disk. 
Possible explanations for the apparent shape of the soft excess 
include continuing 
problems with the softest energy calibration, and the more intriguing 
possibility of a peak due to the presence of a blend of broadened emission lines 
as suggested previously for some Seyfert galaxies 
\citep{br,Turnerea01}. \citet{Comastriea98} present a {\it BeppoSAX} spectrum 
of Ton~S180 down to $\sim 0.1$ keV, with no evidence for a spectral drop
below 0.3 keV, supporting the possibility that this is due to residual 
calibration problems in the LETG. Thus we do not perform detailed fitting 
to this feature.

\section{EUVE}
Archival {\it EUVE} data are available
covering a period which overlaps the {\it Chandra} observation.
We processed these data using standard techniques.
Source counts were summed in a circular aperture of 25
pixels in radius and the background calculated from a surrounding annulus
of 30 pixels in width.
The deadtime-Primbsching correction was used to correct the count rate for the
loss of events due to the deadtime of the detector electronics and
``primbsching'' caused by the rather limited width of the telemetry buffer. A standard
technique calls for data
with a deadtime-Primbsching correction (DPC) factor $>$ 1.25  to be discarded,
as the systematic errors present in the estimates of this correction factor
increase with its magnitude. However, during the course of reducing these data
it was noted that the DS DPC factor frequently lay above 1.5,
significantly greater than the more typically observed values of 1.0-1.3. This
was likely due to increased geocoronal emission possibly associated with the
approaching solar maximum at that epoch and/or the decreasing orbital altitude of
the {\it EUVE} satellite. These conditions forced us to select data
between the more liberal limits of $1.0 <$DPC$< 2.0$ in order to have
adequate counts for construction of a spectrum for the time period
simultaneous with the {\it ASCA} and LETG overlap observation.
The effective on-source exposure was 21 ks within the start
and stop times determined from the LETG observation. The
background-subtracted count rate for the full-band was
$7.1\pm3.7 \times 10^{-3}{\rm cm\ s^{-1}}$.

\section{FUSE}
\subsection{Data Reduction}

We used {\it FUSE} to obtain the 905--1187\,\AA\ far-UV spectrum of 
Ton~S180 on 1999 December 12, 05:50:32--19:41:14 UT.
The total observing time was 15.2\,ks. 
For a full description of {\it FUSE} and its initial in-flight performance, 
see \citet{Moosea00} and \citet{Sahnowea00}.
Briefly, four separate primary mirrors in {\it FUSE} collect light to feed
four prime-focus, Rowland-circle spectrographs.
Two photon-counting micro-channel-plate detectors
with KBr photocathodes image the dispersed light.
Two of the optical systems use LiF coatings and produce spectra
covering the $\sim 990$--1187\,\AA\ wavelength range.
The other two systems use SiC coatings on the optics to provide
spectral coverage down to 905\,\AA.
Our observations of Ton~S180 used the
$30\arcsec \times 30\arcsec$ low-resolution apertures.
We obtained good spectra from the LiF1 and LiF2 channels covering the
987--1187\,\AA\ band, and lower signal-to-noise-ratio (S/N) spectra from the
SiC1 and SiC2 channels covering 905--1091\,\AA.
The flux scale is accurate to 
$\sim$10\,\%, and the wavelength scale
is accurate to $\sim$15\,\kms.
For detailed analysis of line and continuum fluxes, we bin this
spectrum by 5  pixels, preserving the $\sim20$\,\kms{} resolution
of this observation.  
Figure~\ref{fig:lc}  shows that while the {\it FUSE} observation was not 
performed simultaneous with {\it Chandra} it does cover a 
similar mean flux state to that covered by the {\it Chandra} observation. 

\subsection{FUSE Results}

\begin{figure}	 
	\epsscale{0.9} 
	\vspace{-1.6truecm}
	\plotone{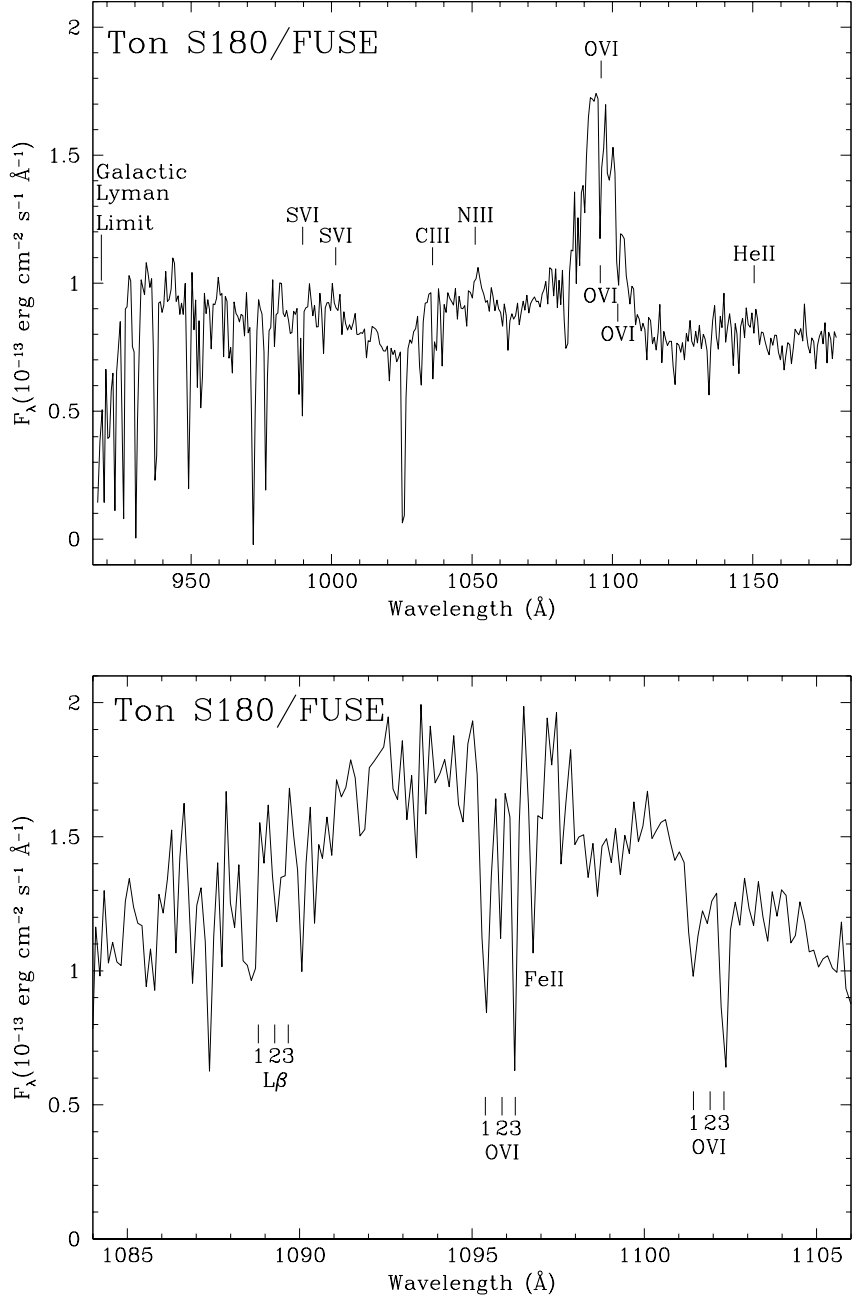}
	\caption[{\it FUSE} and {\it Chandra} versus ASCA]
{Top: {\it FUSE} spectrum of Ton~S180 
binned to a resolution of 0.6\,\AA\ (100 pixels). 
In addition to the noticeable broad \ion{O}{6}
emission line, suggested identifications for other weak far-UV emission lines 
are marked. The associated  \ion{O}{6} absorption lines are indicated.  
All other absorption lines are foreground Galactic or intergalactic features.
Bottom: the section of the {\it FUSE} spectrum surrounding the peak of the
broad \ion{O}{6} emission line is shown. The spectrum is binned to a
resolution of 0.12\,\AA\ (20 pixels) to show the continuum and emission lines 
more clearly. The three associated \ion{O}{6} absorption systems are marked, 
as well as the corresponding locations expected for Ly$\beta$ absorption.
The \ion{Fe}{2} feature is foreground Galactic absorption. 
\label{fig:fuse} 
}
\end{figure}

As seen in Figure~\ref{fig:fuse}, the far-UV spectrum of Ton~S180
shows a bright, blue continuum and prominent broad \ion{O}{6} emission.
Fainter emission from \ion{S}{6} $\lambda\lambda934,945$, 
\ion{C}{3} $\lambda977$, \ion{N}{3} $\lambda991$, 
and \ion{He}{2} $\lambda1085$ may also be present.
The foreground Galactic and intergalactic absorption visible in
this spectrum has already been discussed by \citet{Savageea00}, 
\citet{Sembachea00}, and \citet{Shullea00}, one noteworthy
feature is the deep absorption by Ly$\beta$. 
In addition to these foreground features, absorption at three velocities 
near the redshift of Ton~S180 is visible in the 
\ion{O}{6} $\lambda\lambda$1032,1038 resonance doublet (Figure~\ref{fig:fuse}).

To measure the strengths of these features and that of the broad \ion{O}{6}
emission, we used the IRAF task {\tt specfit} \citep{Kriss94}.
We used a power law for the underlying local continuum, a broad
Gaussian for each of the \ion{O}{6} emission lines with their
fluxes fixed at a 2:1 ratio, and Gaussians for
the narrow absorption lines. The broad \ion{O}{6} lines have a
full-width at half maximum (FWHM) of $2600\pm186$\,\kms{} and a total
flux of $(1.10\pm0.05) \times 10^{-12}$\,\eflux. 
They are blueshifted relative to the systemic redshift 
by $490\pm69$ \kms. 
Unfortunately, the Ly$\beta$ lines corresponding to
these velocities fall in  the gaps between the LiF detector segments, and
so we must use the lower S/N  SiC2A data to measure their strengths.  No
Ly$\beta$ lines are detected. 

For the \ion{O}{6} absorption lines the measured 
equivalent widths ($W_\lambda$) 
column densities and full-widths at half maximum (FWHM) are 
summarized in Table~\ref{FUSEabslines}. 
The doublets have optical depths 
consistent with a 2:1 ratio,  implying full coverage of the
underlying continuum and broad emission lines.
Given the strength of the  \ion{O}{6} absorption and the weakness of any
corresponding neutral hydrogen, we conclude that this gas is in a fairly
high state of ionization.

\section{HST} 

\subsection{STIS Data Reduction}

As {\it HST} was in safe-mode during 1999 December the earliest we could
obtain {\it HST}/STIS observations of Ton~S180 was 2000 January 22 (UT). We
used the 52\arcsec x 0\farcs2 slit to obtain a UV spectrum over the
range 1150--3150\,\AA\ at a spectral resolving power of
$\lambda$/$\Delta\lambda$ $\sim $ 1000. The exposure times were 1260\,s for
the G140L grating and 720\,s for the G230L grating. We reduced the spectra
with the IDL software developed at NASA/GSFC for use by the STIS Instrument
Definition Team \citep{Lindler98}. 
We then combined the G140L and G230L spectra in the region
of overlap for display purposes.

\subsection{STIS Results}

\begin{figure} 
	\epsscale{0.8} 
	\plotone{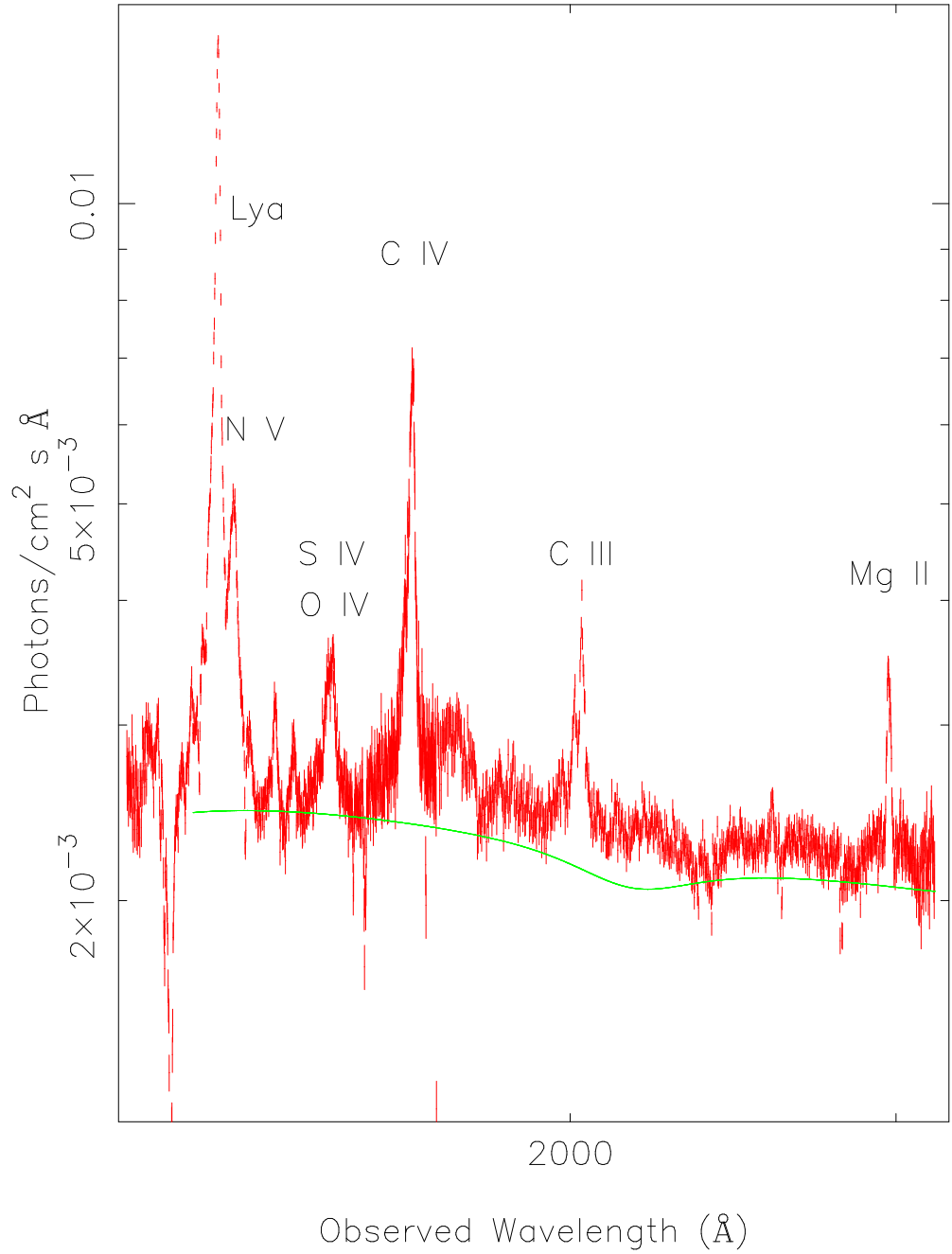} 
	\caption[]{The {\it HST} STIS (1150--3150 \AA) 
data, the solid line shows a power-law of 
energy index $\alpha=0.66$ convolved with extinction by E(B-V)=0.0296,
(i.e. neither the data or model has absorption correction). 
The strong lines are labelled, 
lines are tabulated in detail in Table~\ref{STISemslines}.
\label{fig:fusestisfit}
}
\end{figure}

Figure~\ref{fig:fusestisfit} shows the STIS spectrum of Ton~S180. 
The underlying continuum form is the primary objective of this study, and 
to this end, we first fit simple power-law models to the STIS data, 
absorbed by E(B-V)=0.0296, the extinction due to Galactic material in the 
line-of-sight. The best-fitting power-law has slope $\alpha=0.66\pm0.14$. 
The extrapolation of this continuum 
slope provides a good fit to the {\it FUSE} continuum, after correction 
of the {\it FUSE} data for absorption (using reddening curves from 
\citealt{hutch} and \citealt{clay}). 
However, 
the source continuum level in the STIS data is lower than that observed 
in Dec 1999 by {\it FUSE}.
We find the normalizations of the STIS and 
{\it FUSE} datasets show a flux discrepancy, in the sense that the 
STIS data find the source at 55\% of the flux level observed by {\it FUSE}. 

The uncertainty on absolute flux is $\sim 10\%$ for {\it FUSE} and 
$\sim 2\%$ for STIS data. Thus we attribute the  discrepancy to 
variability in Ton~S180 over the $\sim 5$ weeks separating those
observations. Figure~\ref{fig:lc} shows that the {\it FUSE} 
observation covered a similar flux state to the {\it Chandra} observation, 
so we do not want to rescale the {\it FUSE} data as it samples the same 
flux state as the X-ray data. 
Given an expectation of lags between emission in the different wavelength regimes 
of an AGN, it is always difficult to assess how to construct the most meaningful 
and instructive SED. The flux discrepancy in data from the 
overlaping bandpass of the STIS and {\it FUSE} data 
remove ambiguities as to breaks in intrinsic spectral shape, thus 
for construction of the SED of Ton~S180 we scaled-up the STIS and 
ground-based data by a factor 1.78 to compensate for the flux variability. 

The absorbed powerlaw continuum extrapolates from the STIS band 
to agree with the {\it FUSE} continuum form. 
No spectral turnover is evident in these data, indicating that if the BBB 
component is contributing to the {\it FUSE} data, then its peak lies 
above $\sim 912$ \AA{} (15 eV). 
The excesses above the continuum fit are due to known emission features 
which are detailed in Table~\ref{STISemslines} and 
some weak absorption features are also evident. In addition to 
the distinct lines there is evidence for 
emission from \ion{Fe}{2} and \ion{Fe}{3} between 1900 and 2300\,\AA\, 
similar to that observed in another NLSy1, I ZW 1 \citep{vestwilk01}. 
Examination of the detailed STIS data shows that all of the emission lines  
expected for a Seyfert 1 galaxy are present, as well as a number of absorption
lines from our Galaxy. The UV emission lines are narrow compared to those in
typical Seyfert 1 galaxies; for example, the full-width at half-maximum of the
\ion{C}{4} $\lambda$1549\,\AA\ line is $\sim 2300\pm80$\,\kms. 
The peak of  \ion{C}{4} is
blueshifted by $510\pm20$\,\kms{} with  respect to the systemic velocity.
The emission lines
are somewhat asymmetric, as illustrated by the \ion{C}{4} line, with more
emission in the blue wing than red wing. There is no evidence for intrinsic UV
absorption lines, which occur in $\sim$60\,\% of normal and narrow-line Seyfert 1
galaxies \citep{Crenshawea99}. Although we cannot rule out the possibility of
weak absorption at this resolution, we estimate an upper limit on the equivalent
width of any  \ion{C}{4} absorption to be 0.3\,\AA.

\section{Ground-based Data}

\subsection{$uvby$ Photometry}
Str\"omgren $uvby$ observations \citep{Strom56} were made of Ton~S180 with 
the Danish 1.5m
telescope at ESO, La Silla, on the night of 2000 January 20-21.
Observations of two secondary standards (DM-261339 and DM-38022) were 
also made. Each standard was observed twice in a $uvbyybvu$ sequence 
before and after the single observation, $uvby$, of Ton~S180. 
Each image was pre-reduced in the same standard manner. The bias, 
calculated as the mean of the overscan region, was subtracted 
and then the image was flat-fielded using the adopted sky-flat for 
the observing run. Instrumental magnitudes were extracted from the images 
using DAOPHOT{\sc ii}. Simple aperture photometry is all that is 
required since neither the standard stars nor Ton~S180 reside in crowded 
fields. A 25\farcs0 (32.05 pixel) aperture was adopted. 
The sky background 
was estimated using an annulus with inner and outer radii of 
64\farcs1 (50 pixels) and 70\farcs5 (55 pixels). 
For all the $uvby$ photometry discussed here, atmospheric 
extinction corrections were made based on extinction 
coefficients determined at the Danish-50 cm telescope. On the night 
of these observations the extinction was marginally higher than the mean 
while the rms residuals between observed and catalogue indices 
of standard stars are a factor 3--7 larger than on nights of excellent 
photometric quality. 

Transformations have been determined between instrumental and 
standard systems based upon 170 Str\"omgren $uvby$ measurements of 42 
secondary standard stars (observed in 1999, January, February, November 
and December). 
We transformed the Ton~S180 instrumental photometry to the standard 
system, based upon the December standard transformation relations. 
Correcting for time-evolution of the relation, and for 
the difference in $b$ filter used for the target, versus that used 
for the standard stars 
we obtain the following Str\"omgren magnitudes and indices for 
Ton~S180: $u=14.94\pm0.03$\,mag, $v=15.03\pm0.02$\,mag, 
$b=14.74\pm0.02$\,mag, $y=V=14.58\pm0.02$\,mag, $(b-y)=0.16\pm0.02$\,mag and
$c_1=-0.380\pm0.04$\,mag. To take into account the various possible error 
sources deriving from a) the use of a different $b$ filter and b) the 
zero-point offsets, we have adopted uncertainties of twice the rms 
residuals for the full 1999 standard star dataset. 

We convert $uvby$ magnitudes into monochromatic fluxes via the 
equation $f_{\lambda}=10^{m_{\lambda}/-2.5}F_{\lambda,m_{\lambda}}=0$
where the calibrating fluxes $F_{\lambda,m_{\lambda}}$ are taken from 
\citet{Pritchardea98}.  The fluxes obtained are listed in Table~\ref{uvby}.
The errors on fluxes are propagated from those on the magnitudes.

\subsection{Infrared Photometry}

On 2000 Jan. 23 (JD 2451567.54) we observed Ton~S180 with the
near-IR imager/spectrograph OSIRIS mounted on the CTIO 1.5~m telescope.
A total of 300~s in $K'$, and 150~s in both $J$ and $H$ were
recorded for Ton~S180 along with the standard star 9103 
\citep{Perssonea98}. The standard star is located in the vicinity of Ton~S180 at a
similar airmass.  The instrumental magnitudes were transformed
to the system of \citet{Perssonea98} based on the offsets found for the
standard star. This gave $J=13.22\pm0.04$~mag, $H=12.60\pm0.03$~mag, and
$K_s= 11.67\pm 0.03$~mag where the errors are random errors from photometry
and zero points. The calibration was checked against another standard
observed about an hour later at a similar airmass, yielding good
agreement.

  To convert the magnitudes to monochromatic fluxes we use the $J$ and
$H$ zero magnitude fluxes computed by \citet{Cohenea92} which are based on
Kurucz models of Vega and Sirius. These models are computed for the
UKIRT system \citep{CasaliH92} but \citet{Perssonea98} show
that their magnitudes have a similar zero point. 
For the $K_s$ zero magnitude flux we adopt the value from \citet{Tokunaga97} 
quoted in the OSIRIS users manual. The resulting monochromatic
fluxes, $f_\lambda$, computed as in the previous section are given in
Table~\ref{uvby}. 

\section{Examination of the SED}

\begin{figure}	 
	\vspace{-1.0truecm}
	\epsscale{1.0} 
	\plotone{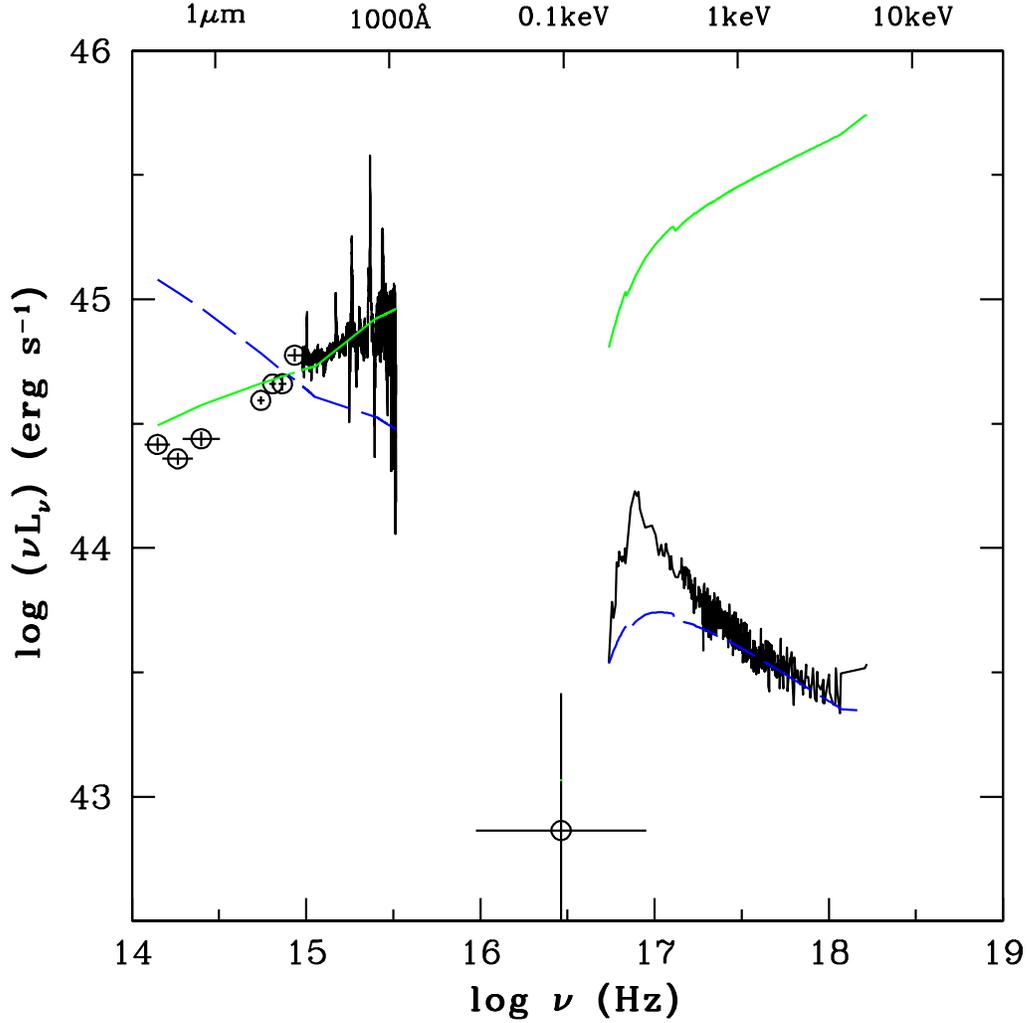}
	\caption[]{The multi-waveband data for Ton~S180. 
In this plot, the data have not been corrected for absorption. The 
blue dashed line shows the (absorbed) power-law $\alpha=1.44$ from 
the 2-10 keV regime, 
extrapolated to lower energies. The green solid line shows the power-law 
$\alpha=0.66$ (convolved with line-of-sight absorption) from the fit to 
the STIS and {\it FUSE} data, extrapolated to
higher energies. The open point between log $\nu=$16--17 Hz 
represents the {\it EUVE} data, 
the circles represent the ground-based data.
\label{fig:pl_extraps}  
}
\end{figure}

First we compared the SED data to the power-law continua determined 
for the UV and X-ray regimes. 
Figure~\ref{fig:pl_extraps} shows the extrapolation of the 
best-fitting power-law to the {\it HST}/STIS data 
($\alpha=0.66$) greatly overpredicts the 
X-ray flux. Clearly the spectrum 
must break somewhere between the UV and soft X-ray regimes. 
Also shown is the hard X-ray continuum slope, $\alpha=1.44$, extrapolated to 
lower energies. This continuum intercepts the UV data around a few thousand 
\AA\, but again, the hard X-ray power-law must terminate somewhere 
between the UV and soft X-ray regimes, as it overpredicts the 
optical and infrared data. 

In order to examine the approximate energy distribution for Ton~S180 
we first corrected the data for the small amount of extinction  
due to the Galactic line-of-sight gas. 
In the STIS band the reddening correction was made following \citet{card89} 
and in the {\it FUSE} band using \citet{hutch} and \citet{clay}, both 
assuming E(B-V)=0.0296, the Galactic extinction. 
The absorption correction in the X-ray regime was made following 
\citet{mm83} and assuming a Galactic value 
$1.55 \times 10^{20} {\rm cm^{-2}}$. 
Table~\ref{sedvalues} summarizes some useful data from the SED. 
A simple 
parameterization was made of the spectral shape using 
the hard X-ray power-law, $\alpha=1.44$ breaking to $\alpha=2.5$ at 1 keV 
and then breaking to $\alpha=0.66$ at 0.1 keV. This parameterization is shown 
as a solid green line in Figure~\ref{fig:sed}. The peak of the SED 
in this case is 80 eV. 
The dotted green lines 
denote the uncertainty in the intrinsic SED, due to some uncertainty 
in the line-of-sight absorption measurement.  
Parameterization of the 
soft X-ray regime as a steep power-law is clearly inadequate, and we also 
overlay an alternative model with continuum plus {\sc diskbb} soft component. 
It is interesting to see that the best-fitting {\sc diskbb} model, 
which has a temperature 
of 98 eV at the inner radius, 
 does not predict any BBB component 
would appear in the UV band. 

Even application of the physically meaningful models such as 
{\sc diskbb} leave some unmodeled structure in the soft component, 
i.e.\ a sharp spectral break below 0.3 keV. It is currently unclear 
whether this structure represents the intrinsic form of 
the soft X-ray emission or whether it represents a residual uncertainty in 
the ACIS/LETG calibration.
The break is not well modeled using  neutral or ionized gas. The most 
obvious possibility remaining is that this sharp feature is due to the presence 
of emission features. However, uncertainty in calibration prompts us to 
note this structure but not to model it in detail.

\section{Discussion}

\subsection{Interband Indices}

Table~\ref{aox} shows the indices between various wavebands for Ton~S180, compared to 
some values previously determined for other Seyfert type galaxies. 
A commonly cited slope is $\alpha_{\rm ox}$, and the value 
$\alpha_{\rm ox}=1.52\pm0.02$ derived for Ton~S180 is consistent with 
$\alpha_{\rm ox}=1.46^{+0.05}_{-0.07}$
found for a sample of optically-selected radio quiet AGNs \citep{zam81}. 
Ton~S180 
appears X-ray weak compared to the mean index of $1.14\pm0.18$ determined 
for the {\it ROSAT} International X-ray/Optical Survey 
(RIXOS, \citealt{puch96})
however, Ton~S180 does lie within the 
the broad range found for RIXOS sources, which include both BLSy1s and NLSy1s. 
The value $\alpha_{\rm ox-hard}$ was 
defined in \citet{gp98} as the index linking 
5500\,\AA\ and 1 keV, and Ton~S180 lies within the 
broad ranges found for soft X-ray 
and hard X-ray selected AGN from that study, based on {\it ROSAT} observations.

There are two questions of interest here, one is 
whether NLSy1s as a class have systematically different interband indices 
to BLSy1s, and the other is whether Ton~S180 is unusual compared to other 
NLSy1s. 
\citet{nag01} have broached the first question by comparing 
the quantity $\alpha_{\rm ox}$ for NLSy1s and BLSy1s. Those authors find 
average values and $1 \sigma$ deviations 
$\alpha_{\rm ox-NLSy1}=1.31\pm0.16$ and 
$\alpha_{\rm ox-BLSy1}=1.36\pm0.24$, thus concluding there  to be  
no significant differences between 
this quantity for the two extremes of the Seyfert 1 population, 
contrary to some previous results \citep[e.g.][]{puch96}. 

In summary, 
based upon the comparison of interband indices with other studies yields 
no evidence that Ton~S180 has an unusual ratio of optical/UV to X-ray flux. 
While opinions in the literature differ on whether there is a systematic difference 
in $\alpha_{\rm ox}$ for the extremes of the Seyfert 1 population, it seems 
clear that interband indices have large ranges and their use is best suited to 
comparison of large samples of sources. In this study we proceed by more 
detailed examination of the shape of the SED, and comparison of our data with 
other detailed SEDs.

\subsection{The Form of the SED}

A SED for Ton~S180 was first presented by \citet{Comastriea98}, 
who found the soft X-ray component to contain the bulk of 
the energy in this Seyfert galaxy. This campaign of observations 
provides a more complete SED for Ton~S180 than previously available, 
with a large amount of simultaneous data. 

\begin{figure}	 
	\epsscale{0.9} 
	\plotone{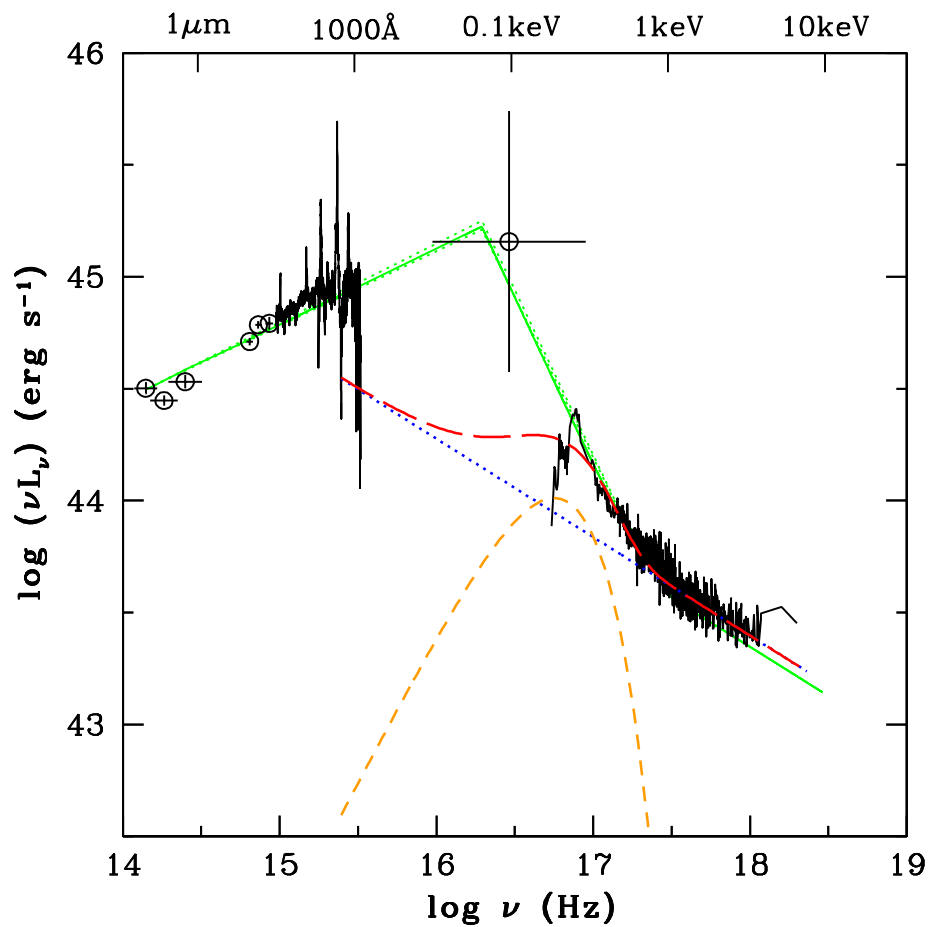} 
	\caption[The Spectral Energy Distribution of Ton~S180]
{The Spectral Energy Distribution of Ton~S180. The data presented here are 
shown as a solid black line. The data have 
been corrected for Galactic line-of-sight extinction. The circles 
represent the ground-based data. 
The simple model parameterization of the SED is shown as a solid green line.
The dotted green line straddling this shows the uncertainty in the SED 
due to uncertainty in the galactic line-of-sight absorption. 
The dashed orange line is the blackbody and the dotted blue line the power-law 
model components of the SED, the red dashed line is their sum. 
\label{fig:sed}
}
\end{figure}

Examination of the detailed energy distribution of Ton~S180 
reveals significant differences compared to some 
other well-studied AGN.
Figure~\ref{fig:sed} shows the extinction-corrected SED 
of Ton~S180, and some parameterizations of its form. 
Overlaid on the parameterizations of the SED of Ton~S180 
are the SEDs of other AGN; NGC~5548 
is shown as a magenta dash-dotted line \citep{KCFP98} while 
the mean radio-loud and radio-quiet 
quasars (from \citealt{Elvisea94}) are shown as dotted black 
and dashed blue lines, respectively. 
The most immediate result is that the SEDs of the Seyfert galaxies appear 
to peak somewhere in the extreme UV/soft X-ray band, while the quasars 
peak in the UV regime. Furthermore, 
the SED of Ton~S180 peaks at a higher energy 
than that of NGC~5548.

Some caution is required in the comparison of SEDs constructed 
with different datasets and various assumptions. 
Some apparent difference in SEDs could be 
an artifact of the assumption of some continuum form for the quasars, 
versus a simple joining of the soft X-ray to UV data for the Seyferts. 
However, such assumptions are only necessary in the problematic regime 
between $\sim 900$\AA\ and $\sim$ 0.1 keV. We find the evidence for 
true underlying 
differences between Ton~S180 and the comparison sources to be strengthened 
by the absence of a contribution from the BBB in the UV band of Ton~S180.

\begin{figure}	 
	\epsscale{0.9} 
	\plotone{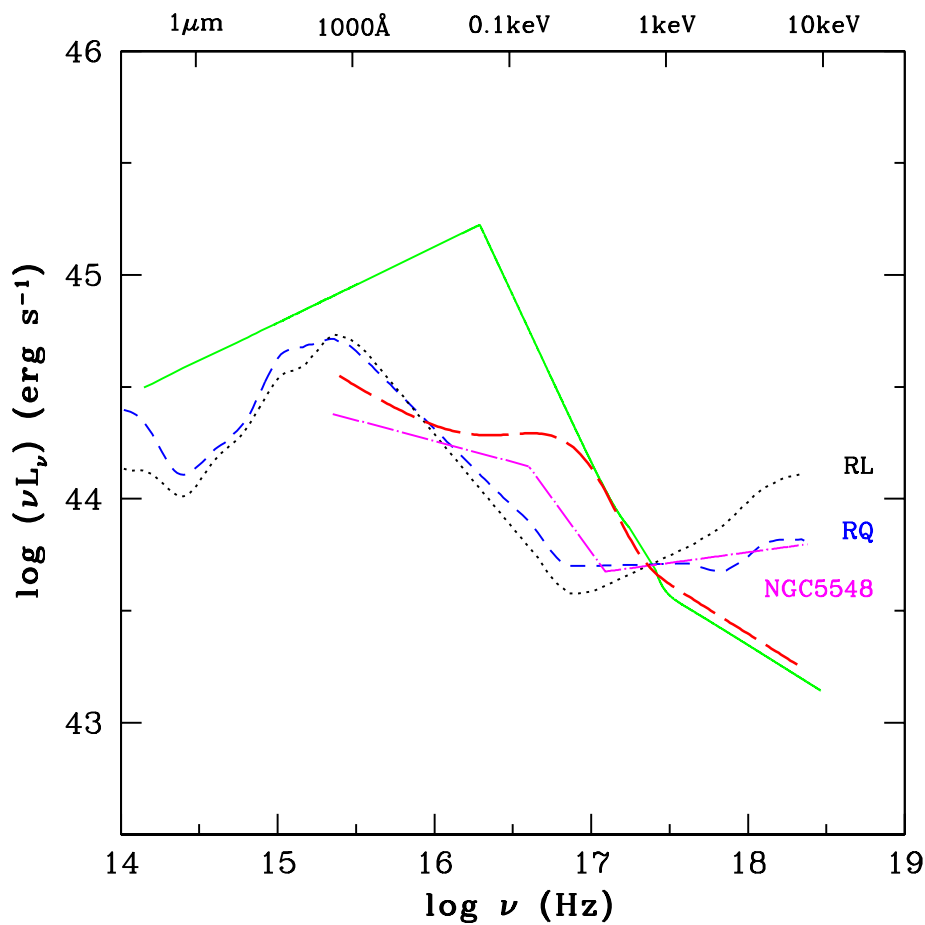} 
	\caption[The Spectral Energy Distribution of Ton~S180]
{The Spectral Energy Distribution of Ton~S180, 
represented by the solid green line from the simple parameterization. 
The dashed red line shows the sum of the blackbody and power-law model 
components (as in Figure~7). 
For comparison, the mean SED for radio-loud and radio-quiet 
quasars are also shown as dotted black and dashed blue lines, respectively 
\citep{Elvisea94}. The SED of NGC~5548 (a Sy1.5) is shown as a 
magenta dash-dot line
\citep{KCFP98}.
\label{fig:sedmo}
}
\end{figure}

A standard optically thick, geometrically thin 
accretion disk \citep{SS73} predicts the temperature of the peak 
of the disk spectrum $T(R)$ to be a function of 
the mass of the central black hole and the accretion rate: 
$T(R) \sim 6.3 \times 10^{5} \; (\dot{M}/\dot{M}_{\rm Edd})^{1/4} \;
        M_{8}^{-1/4} \; (R/R_{\rm S})^{-3/4} {\rm K}$ \citep{Peterson00},
where $R$ is the radius, $R_{\rm S}$ is the Schwarzschild radius, 
$M_8$ is the mass in units of $10^8 \Msun$ and $\dot{M}$ is the 
accretion rate in units of the Eddington accretion rate. 
All other things being equal, a difference of two orders of magnitude in mass 
should yield a disk spectrum whose peak energy is a factor of 3 lower for the 
higher-mass system than the lower-mass system; similarly, 
a factor 100 increase in  accretion 
rate (relative to the Eddington rate) 
shifts the peak to a factor 3 higher energy. 
Thus the RQQ SED (Figure~\ref{fig:sedmo}) represents 
sources radiating at substantially below  
the Eddington accretion rate, and having a high central mass; these 
yield a relatively cool disk spectrum. At the other extreme, NLSy1s 
are thought to be accreting close to the Eddington limit, and have 
a low mass; the disk spectrum appears hot. The BLSy1 NGC~5548 
represents an intermediate system in terms both of the accretion rate and 
central mass, and this appears to have an intermediate SED.
\citet{Telfer02} find that for a sample 
of QSOs, the entire continuum from 10 eV to 2 keV can be represented by a 
single power-law; this is clearly not the case in Ton~S180 where the 
X-ray spectrum steepens below 2 keV, and neither the soft or hard X-ray 
components extrapolate to meet the UV data 
in a satisfactory way\footnote{While the {\it Chandra} data 
show a turnover at 
0.3 keV, this turnover is sharper than that expected from observation of 
the peak of the disk spectrum and 
there is some discrepancy between these data and data from other instruments.
 For these reasons,  henceforth we will assume this turnover 
does not indicate the peak temperature of the disk.}.

NGC~5548 has a central black hole mass estimated at 
$\sim 10^8 \Msun$ \citep{K2000}. 
If the peak of the SED for NGC~5548 is close to  
the Lyman limit ($T \sim 1.6 \times 10^5$K) 
then an accretion rate of 11\% of Eddington would be  
estimated, assuming a standard thin disk picture. 

Few strong constraints exist on the central mass in Ton~S180. 
From variability observed in the X-ray regime 
\citet{Romanoea02} found
$M_{\rm BH} \gtsim 8 \times 10^{6} \Msun$ for Ton~S180; however,
those authors assumed a bolometric luminosity which is lower than
that revealed by this SED, leading to a revised limit
$M_{\rm BH} \gtsim 8 \times 10^{7} M_{\odot}$. Mass estimates
such as these can be misleading if the X-ray variability is due,
for example, to flares on the surface of the accretion disk, as
the timescale of variation may not be directly related to the
scale-size of the disk system. Thus we examine an
alternative estimate of mass based upon the
luminosity at
5100 \AA\ ($ \nu L_{\nu} \sim 3 \times 10^{44}{\rm erg\ s^{-1}}$).
Using the relation derived from other NLSy1s
(\citealt{Peterson00}, their Figure~7) we estimate
a central mass $M \sim 2 \times 10^7  \Msun$
(with a factor $\sim 2$ uncertainty) and the broad line region 
to exist at a radius $\sim 100$ light days 
(\citealt{Peterson00}, their Figure~6). This is in keeping 
with the systematically large BLR radii suggested  
by  \citet{gianstirpe} for NLSy1s compared to BLSy1s. 
As the level of starlight contamination of the 
5100 \AA\ flux is difficult to assess, this mass should be considered 
as an upper limit on the true central mass. 
The difference in SED peak energies is thus expected, as 
Ton~S180 has a lower mass than NGC~5548, and NLSy1s are thought to have 
systematically higher accretion rates than BLSy1s. 

Unfortunately, the peak of the spectrum in Ton~S180 remains 
loosely constrained. The {\it EUVE} data favor the simple parameterization 
of the XUV spectrum (Figure~\ref{fig:sed})  
indicating that the peak lies close to or below 100 eV. 
Assuming a standard disk spectrum, the disk temperature must 
be greater than $\sim 15$\,eV, 
the peak of any cooler component of significant flux would 
show up in the {\it FUSE} data. The SED data indicate 
a peak close to 100 eV. 
Assuming a peak in disk emission for Ton~S180 at this energy,
which corresponds to $4 \times 10^5$\,K, then for 
$M \sim 2 \times 10^7  \Msun$, 
$\dot{M} \sim 0.88 \times \dot{M}_{\rm Edd}$. 
For black holes operating near 
the Eddington limit the accretion disk surface is predicted 
to be highly ionized.   
There is certainly strong evidence for an ionized disk in Ton~S180, as 
the Fe K$\alpha$ line appears to be produced in highly ionized material
in {\it BeppoSAX} \citep{Comastriea98} and {\it ASCA} data \citep{TGN98}. 

The results from Ton~S180 appear to fit into the standard 
disk picture. However, we also note that 
 \citet{CGK91} conclude that the standard disk model 
 is not applicable to the UV spectra of quasars. Their 
 case rests on the lack of any relation 
 between $\alpha_{\rm uv}$ and luminosity. 
However, since the peak of the disk spectrum is generally at rest
wavelengths of 1000 \AA\ or shorter \citep{Zhengea97,Telfer02}, 
$\alpha_{uv}$ is indicative of only the rising edge of
the disk spectrum.  In the standard disk model, the spectral slope
in this region is relatively insensitive to luminosity, so one
does not necessarily expect a strong correlation between $\alpha_{uv}$
and luminosity of the BBB. 

As a final note, the lowest frequency IR point lies above the
adjacent IR points. This has been observed in many AGN and 
is due to thermal emission from dust grains heated to
close to their evaporation temperature (1500 K for graphite)
close to the central engine \citep{rieke78}.  
Recently, strong near-IR emission from the Seyfert 1 galaxy
NGC 7469 has been attributed 
to very hot dust grains ($T > 900$ K) associated with the putative torus 
\citep{marco}, 
this is also observed in the SED of NGC3783 \citep{alloinea95}.

\subsection{The Energy Budget of Ton~S180}

The multi-power-law parameterization of the SED for Ton~S180 
makes it possible to estimate the luminosity in various 
energy regimes, helping to constrain 
reprocessing mechanisms and isolate the primary energy source.
Table~\ref{enbudget} shows the observed and intrinsic luminosities in several 
energy-bands, defined in the rest-frame of the 
source. The implied bolometric luminosity is 
$L_{bol} \sim 10^{46}\,{\rm erg\ s^{-1}}$. 
More luminosity emerges in the 10 - 100 eV regime than the 
100 eV to 10 keV regime. Assuming that we are seeing all emitted radiation 
in each wavelength regime then this indicates that the 
EUVE--soft X-ray  band contains the 
primary spectral component, in keeping with disk-corona models 
(e.g. \citealt{hm91}). 

The energy budget and the SED show that Ton~S180 is 
relatively X-ray weak above $\sim 2$ keV (interband 
indices are insensitive to this, as historically X-ray fluxes for comparison 
with optical fluxes have been taken at soft X-ray energies). 
One possible reason for this is Compton-cooling of the hard spectrum 
by the large flux of soft X-ray and UV photons, as discussed by many authors, 
including \citet{PDO95}. 

\subsection{The State of the Circumnuclear Gas}

The weak \ion{O}{6} absorption features detected in the {\it FUSE}\  data,
the absence of absorption from lower ionization species in the {\it HST}
data, and the lack of detectable X-ray absorption in the {\it Chandra}
spectrum together indicate the presence of a small column of
circumnuclear material which appears to be in a high state of ionization
compared to that observed in other well-studied sources such as NGC~5548
and NGC~3783. The outflow velocity of $\sim 500\ {\rm km\ s^{-1}}$ is not
unusual. Many Seyfert galaxies have shown evidence for outflow in UV and
optical data (e.g. \citealt{Cren97}). In the X-ray regime {\it Chandra}
grating observations have revealed supporting evidence for outflowing
gas, with velocities of order a few hundred ${\rm km\ s^{-1}}$ (e.g.
\citealt{Collea01,Kaastraea00,K2000,K2001}). A picture of Ton~S180 being
shrouded by highly ionized gas is consistent with earlier BeppoSAX
\citep{Comastriea98} and {\it ASCA} \citep{TGN98} observations of Fe
K$\alpha$ emission, as well as the new  {\it ASCA} data which show that
the narrow component of Fe\,K$\alpha$ is consistent with emission from
\ion{Fe}{25}--\ion{Fe}{26}.

The ratio of the \ion{O}{6} to \ion{H}{1} absorbing columns in the UV
regime is comparable to that of the high-ionization component detected in
Mrk~509 \citep{Krissea00}, which was tentatively identified with the
X-ray warm absorber in that object. However, in Ton~S180 the total
equivalent column density of hydrogen associated with the UV absorber
must be $< 10^{17}~\rm cm^{-2}$; too low to produce detectable X-ray
absorption. However, the absence of ionized circumnuclear gas does not
appear to be a general property of NLSy1s. Some NLSy1s do appear to show
signatures of a warm absorber in the X-ray regime \citep[e.g.
][]{Lee01,Collea01} as well as UV absorption systems
\citep{Crenshawea99}.

To examine the relation between the ionizing spectrum in Ton~S180 and the
circumnuclear gas, we took two estimates of the SED and total ionizing
flux. The conservative estimate links the softest X-ray point at $\sim
0.3$ keV to the highest end of the UV data with a simple power-law. 
For this SED, the total luminosity from
0.01 - 10 keV is $\sim 2.0 \times 10^{45}$ ergs s$^{-1}$ and the
corresponding luminosity in ionizing photons is Q $ \sim 2.9 \times
10^{55}$ photons s$^{-1}$. Taking
instead the (extreme) SED that peaks in the EUV (green line in Figure 7), 
the 0.01 -- 10 keV luminosity is 
$\sim 3.8 \times  10^{45}$ erg s$^{-1}$  
and Q\,$\sim 4.5 \times  10^{55}$ photons s$^{-1}$.

Assuming the typical density and ionization parameter 
in the optical broad-line region (BLR) clouds, 
the radii at which the BLR exists can be estimated.
\citet{WPM99} used these ``photoionization radii'' and the measured
line widths to derive black hole masses, which were in general agreement
with those determined via reverberation mapping. In order to explore the
role of luminosity on the line widths we instead use the masses derived
from reverberation mapping and the photoionization radii ($r$)
to estimate the line widths. Following \citet{WPM99}, we
assume 
${\rm Q}/(4 \pi r^2 c\ {\rm n}_{e} ) \times {\rm n}_{e} \sim 10^{10}$ 
for the line emitting gas.
Based on our estimates of the ionizing luminosity of the central
source in Ton~S180, we derive representative radial distances of r
$\sim 8.8 \times 10^{16}$ cm and $\sim 1.1 \times 10^{17}$ cm, for the
conservative and extreme cases, respectively ($\sim 40$ light days, 
somewhat smaller than the radius estimated from \citealt{Peterson00}). 
If the
BLR clouds are virialized around the central black hole, the FWHM of the
emission lines should be roughly equal to $\sqrt{\rm GM/r}$; for a black
hole mass of 10$^{7} \Msun$, FWHM $\sim$ 1290 km s$^{-1}$ (conservative
case) and 1160 km s$^{-1}$ (extreme case), in reasonable
agreement with the observed FWHM of H$\beta$.

We have estimated the total ionizing flux, Q, for two BLSy1's: for NGC
4151, Q is $\sim$ 2--8 $\times  10^{53}$ photons s$^{-1}$ \citep{K01};
for NGC 5548, Q is $\sim 1 \times  10^{54}$ photons s$^{-1}$
\citep{KCFP98}. Assuming the black hole masses quoted by \citet{WPM99} 
(1.2 x 10$^{7} \Msun$ for NGC 4151, and 6.8 x 10$^{7} \Msun$ for
NGC 5548), the corresponding ``typical'' BLR cloud distances and
velocities are r $\sim 1.2 \times 10^{16}$ cm and 
FWHM $\sim$ 3700 km s$^{-1}$ for NGC 4151, and r $\sim 1.6 \times 10^{16}$ cm 
and FWHM $\sim$ 7460 km s$^{-1}$ for NGC 5548; the FHWMs are in rough 
agreement with the observed values. Clearly, the narrowness of the emission 
lines in Ton~S180 is
partially due to its stronger ionizing flux. Furthermore, given these
``typical'' BLR conditions, it is likely that
the clouds within a few light days of the central source are too
highly ionized even to produce much \ion{C}{4} emission (for a discussion of
range in conditions in which emission lines form, see
\citealt{Baldwinea95}). Hence, the higher value of $Q$ for Ton~S180 requires 
that the BLR gas
is either more highly ionized than in BLSy1s or, if the emission-line
ratios are the same, it must be denser (i.e., denser gas is now ionized
enough to contribute significantly to the emission-line spectrum). Either
way, conditions are not identical to those in typical BLSy1s.

\section{Summary}

Construction of the spectral energy distribution for the bright 
NLSy1 galaxy Ton~S180 shows that most of the energy is emitted in the 
10 -- 100 eV regime, indicating that the primary source of emission 
dominates that band.  The UV and X-ray data together constrain the 
peak of any BBB component to lie 
between 15 and 100 eV. This, and the 
overall shape of the SED indicate that emission from the accretion disk peaks 
at significantly higher energies in this source than in 
BLSy1s, as expected if NLSy1s have smaller central black holes and higher 
accretion rates. 
High-resolution spectra from {\it FUSE}\ reveal UV absorption due 
to \ion{O}{6}. The absence of absorption features in the {\it HST} data 
and the lack of neutral hydrogen absorption in the {\it FUSE} spectrum 
indicate a high-ionization state for the absorbing gas, 
while the absence of soft X-ray absorption that shows that the
column density is quite low.  
The highly-ionized state of the circumnuclear gas 
is most likely linked to the high luminosity and steep slope of 
the ionizing continuum in Ton~S180. 
Given our constraints on the SED in Ton~S180, we find that 
typical BLR emission lines would form at a radius 
which is an order of magnitude further out than in 
typical BLSy1s. The BLR is estimated to exist at a radius 
$\sim 10^{17}$ cm, or 40 light days.

\section{Acknowledgements}

We are grateful to the satellite operation teams for co-ordination 
of the multi-waveband observations. 
This research has made use of the NASA/IPAC Extragalactic database,
which is operated by the Jet Propulsion Laboratory, Caltech, under
contract with NASA; of the Simbad database, operated at CDS, Strasbourg, France; 
and data obtained through the HEASARC on-line service, provided by NASA/GSFC.
This work is based in part on data obtained for the Guaranteed Time Team by the
NASA-CNES-CSA {\it FUSE} mission operated by the Johns Hopkins University.
Financial support to U. S. participants has been provided by NASA contract NAS5-32985.
T.J.\ Turner, G.\ Kriss, S.\ Mathur and P.\ Romano acknowledge support from NASA 
grants  NAG5-7538, NAGW-4443, NAG5-8913 and NAG5-9346 respectively.  
This research was supported by the Danish Natural Science Research 
Council through its Centre for Ground-Based Observational Astronomy. 
Funding for the OSIRIS instrument was provided by grants from The Ohio State
University, and National Science Foundation Grants AST-9016112 and
AST-9218449. The 1.5~m telescope at CTIO is operated by the Association of
Universities for Research in Astronomy Inc.\ (AURA), under a cooperative
agreement with the NSF as part of the National Optical Astronomy
Observatories (NOAO). We thank the anonymous referee for useful comments.

\clearpage  
\begin{deluxetable}{llllc}	
 \tablewidth{0pc} 
 \tablecaption{Observing Log for Ton~S180.\label{obslogs}}
\tablehead{\colhead{Observatory/Telescope} & \colhead{Instrument} & \colhead{UT Dates} 
				       & \colhead{Notes} & \colhead{References} \\
	   \colhead{(1)} & \colhead{(2)} & \colhead{(3)} & \colhead{(4)} & \colhead{(5)}}
\startdata 
{\it ASCA}     	&		& 1999 Dec 03--15     & continuous\tablenotemark{a} & 1,2 \\
{\it Chandra}  	& LETG 		& 1999 Dec 14--15     & continuous       	    & 3 \\
{\it RXTE}\tablenotemark{b}     & PCA		& 1999 Nov 12--Dec 15 & once every 96 min  	    & 2 \\
{\it EUVE}     	&		& 1999 Nov 12--Dec 15 & continuous\tablenotemark{a,c} & 2 \\
{\it FUSE}     	& 		& 1999 Dec 12         & 15.2\,ks; 30\arcsec x 30\arcsec (LWRS) & 4 \\
{\it HST}       & STIS		& 2000 Jan 22         & 6\,ks; 52\arcsec x 0\farcs2 & 4 \\
ESO 1.5m 	& 		& 2000 Jan 21 	      & $uvby$  		    & 4 \\
CTIO 1.5m	& OSIRIS 	& 2000 Jan 23 	      & $JHKs$ 			    & 4 
\enddata 
\tablenotetext{a}{except for gaps due to Earth 
occultation and passage of the spacecraft through the SAA.}
\tablenotetext{b}{Those data are not used in the construction 
	of the SED, but were taken for the complementary 
	timing project \citep{Edelsonea02}.}
\tablenotetext{c}{A subset of these data were used in construction of the SED.} 
\tablerefs{(1)  \citealt{Romanoea02}. (2) \citealt{Edelsonea02}; 
(3) \citealt{Turnerea01}. (4)  This work. }
\end{deluxetable} 


\begin{deluxetable}{lccccc}	
\tablewidth{0pc} 
\tablecaption{UV Absorption Lines\label{FUSEabslines}.}
\tablehead{\colhead{Feature} & \colhead{\#} & \colhead{$W_\lambda$} &
	      \colhead{$N_{\rm ion}$} & \colhead{$\Delta v\tablenotemark{a}$} &
	      \colhead{FWHM}  \\
	  \colhead{} &  \colhead{}    & \colhead{(\AA)}   &
		\colhead{(cm$^{-2}$)} & \colhead{(\kms)} & \colhead{(\kms)} \\
}
\startdata 
\ion{O}{6}  $\lambda$ 1031.93 	& 1 	& $0.20 \pm 0.03$  
	& $1.7\pm 0.22 \times 10^{14}$ & $-146\pm 15$ & $75 \pm 11$  \\
		 		& 2 	& $0.07 \pm 0.001$ 
	& $6.0\pm 0.12 \times 10^{13}$ &   $-6\pm 15$ & $27 \pm 6$  \\
                   		& 3 & $0.21 \pm 0.02$ 
	& $2.0\pm 0.22 \times 10^{14}$ & $+109\pm 15$ & $44 \pm 5$  
\enddata 
\tablenotetext{a}{Velocity is relative to a systemic redshift of $z = 0.06198$.}
\end{deluxetable}

\begin{deluxetable}{lcc}	
\tablewidth{0pc} 
\tablecaption{STIS Emission Lines\label{STISemslines}.}
\tablehead{\colhead{Feature} & \colhead{Flux} &
	      \colhead{FWHM}  \\
	   \colhead{(\AA)}   &
   \colhead{} & \colhead{(km s$^{-1}$)} \\ 
	  }
\startdata
Ly$\alpha$  $\lambda$ 1216 & $30.8\pm6.20$ & $1961\pm122$ \\
\ion{N}{5}  $\lambda$ 1240 & $7.11\pm1.40$ & $1700\pm848$ \\
\ion{Si}{2} $\lambda$ 1260 & $0.49\pm0.19$ &  \nodata \\
\ion{O}{1}   $\lambda$ 1302  & $1.86\pm0.45$ & $921\pm230$ \\
\ion{C}{2} $\lambda$ 1335 & $0.87\pm0.30$ & $1123\pm280$ \\
\ion{S}{4} / \ion{O}{4} \verb+]+ $\lambda$ 1400 & $4.47\pm0.94$ & \nodata \\
\ion{N}{4}\verb+]+ $\lambda$ 1486 & $<0.1$ & \nodata \\
\ion{C}{4} $\lambda$ 1550 & $8.31\pm0.61$ & $2300\pm80$ \\
\ion{He}{2} $\lambda$ 1640 & $1.95\pm0.78$ & \nodata \\
\ion{O}{3} \verb+]+ $\lambda$ 1663 & $0.60\pm0.24$ & \nodata \\
\ion{N}{3}\verb+]+ $\lambda$ 1750 & $0.45\pm0.22$ & \nodata \\
\ion{Si}{3}\verb+]+ $\lambda$ 1892 & $1.25\pm0.25$ & \nodata \\
\ion{C}{3} \verb+]+ $\lambda$ 1909 & $2.16\pm0.43$ & \nodata \\
\verb+[+ \ion{Ne}{4}\verb+]+ $\lambda$ 2324 & $0.24\pm0.10$ & \nodata \\
\ion{Mg}{2} $\lambda$ 2800 & $1.21\pm0.24$ & $1071\pm428$ \\
\enddata
\tablenotetext{a}{Observed fluxes with no absorption correction, 
in units $10^{-13} {\rm erg\ cm^{-2}\ s^{-1}}$. }
\end{deluxetable}


\begin{deluxetable}{lccc}	
\tablewidth{0pc}
\tablecaption{$uvby$ Photometry\label{uvby}.}    
\tablehead{\colhead{Filter} & \colhead{$\lambda$} & \colhead{Mag.\tablenotemark{a}} 
	& \colhead{$F_{\lambda}$\tablenotemark{b}}  \\
	   \colhead{} & \colhead{\AA} & \colhead{} & 
	\colhead{(10$^{-19}$ erg cm$^{-2}$ s$^{-1}$ \AA$^{-1}$)}  
}
\startdata   
$u$      & 3500 &  $14.94\pm 0.03$ & $39.75 \pm 1.02$ 	\\
$v$      & 4110 &  $15.03\pm0.02$ & $ 11.94\pm 0.33$ 	\\
$b$      & 4670 &  $14.74\pm0.02$ & $ 7.71\pm 0.13$ 	\\
$y$      &  5470 & $14.58\pm0.02$ & $ 5.26\pm 0.09$  		
\enddata
\tablenotetext{a}{Str\"omgren magnitude \citep{Strom56}.}
\tablenotetext{b}{No rescaling applied for source variability. }
\end{deluxetable}  


\begin{deluxetable}{lccc}	
\tablewidth{0pc}
\tablecaption{$J$, $H$ and $Ks$ Photometry\label{JHKKs}.}    
\tablehead{\colhead{Filter} & \colhead{$\lambda$} & \colhead{Mag.} & 
	   \colhead{$F_{\lambda}$\tablenotemark{a}} \\
	   \colhead{} & \colhead{\AA} & 
		\colhead{} & \colhead{(10$^{-19}$ erg cm$^{-2}$ s$^{-1}$ \AA$^{-1}$)} 
}
\startdata   
$J$      &  12150  	&   $13.22\pm0.04$  &  	 $3.45 \pm 0.10$	\\
$H$      &  16540  	&   $12.60\pm0.03$  &    $1.40 \pm 0.03$	\\
$Ks$     &  21570  	&   $11.67\pm0.03$  &    $1.14   \pm 0.03$ 	
\enddata
\tablenotetext{a}{No rescaling applied for source variability. }
\end{deluxetable}  

\clearpage

\begin{deluxetable}{lcc}	
\tablewidth{0pc}
\tablecaption{\label{sedvalues}}    
\tablehead{\colhead{Rest Wavelength/Energy}  & \colhead{$\nu L_{\nu}$ (Observed)\tablenotemark{a}}  
	& \colhead{$\nu L_{\nu}$ (Intrinsic)\tablenotemark{a}} \\
	   \colhead{(keV)} & \colhead{}  & \colhead{} \\ 
}
\startdata 
2\,$\mu$m                 &  3.146                   &  3.175 \\
1\,$\mu$m                 &  3.842                   &  3.953 \\
7000\,{\rm \AA}           &  4.283                   &  4.544 \\
5500\,{\rm \AA}           &  4.485                  &  4.823 \\
3000\,{\rm \AA}           &  5.272                  &  6.124 \\
2500\,{\rm \AA}           &  5.356                  &  6.371 \\
1000\,{\rm \AA}           &  6.053                  &  8.892 \\
0.25\,keV                 &  1.521                  &  3.180 \\
1\,keV                    &  0.655                 &  0.686 \\
2\,keV                    &  0.307                &  0.313 \\
10\,keV                   &  0.156                &  0.156   
\enddata
\tablenotetext{a}{In units of $10^{44}$\,erg\,s$^{-1}$. 
$H_0=75$ ${\rm km\ s ^{-1}\ Mpc^{-1},}$  $q_0=0.5$.}
\end{deluxetable}

\begin{deluxetable}{lccccc}	
\rotate
\tablewidth{0pc} 
\tablecaption{Spectral Indices\label{aox}}     
\tablehead{\colhead{Index}  & \colhead{} & \colhead{Definition} &  
	\colhead{Observed}  & \colhead{Intrinsic} & \colhead{BLSy1} \\ 
	 \colhead{} & \colhead{} & \colhead{} & \colhead{}  & \colhead{} & \colhead{}   
} 
\startdata
$\alpha_{\rm 3000\,\AA-1000\,\AA}$      & $\alpha_{\rm uv}$
        & -2.096 log($F_{\rm 1000\,\AA}/F_{\rm 3000\,\AA})$
        & $0.53\pm0.14$ & $0.66\pm0.14$ & $0.85\pm$0.06\tablenotemark{a} \\
$\alpha_{\rm 5500\,\AA-0.25\,keV}$        &
        & -0.489 log($F_{\rm 0.25\,keV}/F_{\rm 5500\,\AA})$
                        & $1.24\pm0.03$  &  $1.12\pm0.03$ & 0.73\tablenotemark{b} \\
$\alpha_{\rm 5500\,\AA-1\,keV}$\tablenotemark{c} & $\alpha_{\rm ox-hard}$
        & -0.378 log($F_{\rm 1\,keV}/F_{\rm 5500\,\AA})$
                        & $1.37\pm0.02$  &  $1.38\pm0.02$
        & 1.13 \tablenotemark{c} \\
$\alpha_{\rm 1\mu m -2\,keV}$           & $\alpha_{\rm ix}$
        & -0.312 log($F_{\rm 2\,keV}/F_{\rm 1\mu})$
        & $1.35\pm0.02$  &  $1.35\pm0.02$  & 1.14-2.16 \tablenotemark{d} \\
$\alpha_{\rm 2500\,\AA-2\,keV}$         & $\alpha_{\rm ox}$
        & -0.384 log($F_{\rm 2\,keV}/F_{\rm 2500\,\AA})$
        & $1.50\pm0.02$  &  $1.52\pm0.02$ &
        $1.46^{+0.05}_{-0.07}$ \tablenotemark{e} 
        $ 1.21\pm0.02$ \tablenotemark{f} \\
$\alpha_{x}$                &
        & -1.431 log($F_{\rm 2\,keV}/F_{\rm 10\,keV})$
        & $1.44\pm0.07$  &  $1.44\pm0.07$   & 0.91 \tablenotemark{g}
\enddata                   
\tablenotetext{a}{ Index in the $\sim$ 2200-1200 \AA\ band \citep{CGK91} } 
\tablenotetext{b}{ \citealt{tea99c}. } 
\tablenotetext{c}{ \citealt{gp98}. } 
\tablenotetext{d}{ \citealt{law97}. } 
\tablenotetext{e}{ \citealt{zam81}. } 
\tablenotetext{f}{ \citealt{puch96}. } 
\tablenotetext{g}{ \citealt{Nandraea97b}. } 
\end{deluxetable}

\begin{deluxetable}{lcc}	
\tablewidth{0pc}
\tablecaption{Luminosities\label{enbudget}}    
\tablehead{\colhead{Energy}  & \colhead{$L_{\rm observed}\tablenotemark{a}$}  
		& \colhead{$L_{\rm Intrinsic}\tablenotemark{a}$} \\
	   \colhead{(keV)} & \colhead{}  & \colhead{} \\ 
	   \colhead{(1)} & \colhead{(2)} & \colhead{(3)} 
}
\startdata   
5$\times 10^{-4}$--0.01       	& 13.94   &  15.69  \\
0.01--0.1        		& 2.86   &  28.80  \\
0.1--1           		& 2.03   &  8.92  \\
1--10           		& 0.65  &  0.66  \\
5$\times 10^{-4}$--10         	& 19.49  &  54.07  \\
0.1--10          		& 2.69   &  9.57  
\enddata
\tablenotetext{a}{In units of $10^{44}$\,erg\,s$^{-1}$.}
\end{deluxetable}

\clearpage

\clearpage 

\clearpage

\clearpage

\clearpage

\clearpage

\end{document}